# Graphene edge from A to Z—and the origins of nanotube chirality

Yuanyue Liu, Alex Dobrinsky, and Boris I. Yakobson

Department of Mechanical Engineering & Materials Science, Department of Chemistry,
and the Smalley Institute for Nanoscale Science and Technology,
Rice University, Houston, TX 77005, USA

The energy of arbitrary graphene edge is derived in analytical form. It contains a
"chemical phase shift", determined by the chemical conditions at the edge. Direct
atomistic computations support the universal nature of the relationship. Definitive
for graphene formation, shapes of the voids or ribbons, this has further important
implications for nanotube chirality selection and control by chemical means, at
the nucleation stage.

An old view that carbons are awkward and intractable to study[1] has changed with
discovery of fullerenes and nanotubes.[2] Recently isolated atomically thin graphite—graphene—
has ignited interest due to both fundamental physics and the hopes for applications.[3,4] While the
lattice of graphene is very strong, significant variability at its edges[5,6] defines the electronic
properties[7,8] as well as the growth dynamics,[9,10] similar to the growth of its close sibling,
nanotubes.[11-13] Motivated by the challenge of possible selectivity, here we derive the graphene
edge energy $\gamma$, from armchair (A) to zigzag (Z) and all intermediate orientation chiral angles, $\chi$.
Supported by the first principles computations, the essential dependence is always a sinusoid,
$\gamma(\chi) \sim \cos(\chi + C)$, but its "chemical phase-shift" C varies with the conditions. This determines
the variation in equilibrium shape of graphene isles or ribbons. Moreover, it has profound
implications in the context of nanotube growth, offering rational ways to control their chiral
symmetry, a tantalizing yet so far elusive goal.

Edge or surface energies both quantify the disruption of interatomic bonds. If all dangling bonds were equal in graphene, the edge energy proportional to their density would be higher for the more tightly packed armchair than for the less dense zigzag, by exactly a factor of $2/\sqrt{3} = 1.15$, Fig. 1a-b. However, this very difference in spacing allows the armchair atoms A to form triple bonds and thus lower their energy relative to the zigzag Z, $\varepsilon_A < \varepsilon_Z$.[11,14] This delicate competition of the energy per atom and their density makes the overall energy balance non-trivial and sensitive to the chemical conditions at the edge.

To derive an analytical expression for the edge energy, we begin with a simple observation that any lattice cut exposes two distinctly different types of atoms: having another edge-atom neighbor, as in a purely armchair edge, or bonded to the 3-coordinated bulk-lattice neighbors, as in a purely zigzag. In Fig. 1a, b and c the computed charge density maps for pure A, pure Z, and a generic chiral edge, show this distinction clearly and support the energy-decomposition ansatz: An arbitrary edge energy can be evaluated as $(C_A\varepsilon_A + C_Z\varepsilon_Z)$ by counting the edge carbon atoms.

With a basis in a honeycomb lattice, an arbitrary edge direction can be specified by two components (n, m), or by the angle between the edge line and the zigzag atomic motif, $\chi$ (to keep with tradition of the chiral angle for nanotubes[2,15]). Inspection of the Fig. 1d then reveals 2m of A-atoms and (n−m) of Z-atoms, over the edge span of $(n^2+nm+m^2)^{1/2}$, henceforth using the lattice parameter $l = 2.46$ Å as a unit. An elementary law of sines, applied to the triangles in Fig. 1d, yields $c_A = (4/\sqrt{3})\sin(\chi)$ for A-type, $c_Z = 2\sin(30°-\chi)$ for Z-type, and $c = (2/\sqrt{3})\cos(30°-\chi)$ for the total edge-atom densities. Adding these, with the appropriate weights $\varepsilon'_A$ and $\varepsilon'_Z$, one obtains the edge energy as $[(4/\sqrt{3})\varepsilon'_A\sin(\chi) + 2\varepsilon'_Z\sin(30°-\chi)]$ per unit length, or

$$\gamma'(\chi) \;=\; 2\gamma'_A\sin(\chi) + 2\gamma'_Z\sin(30°-\chi) \;=\; |\gamma'|\cos(\chi + C') \qquad (1)$$

The last identity makes it clear that the energy must universally depend on the edge direction as sinusoid, with the phase-shift constant determined by the basic edges only, $C' = \arctan(\sqrt{3} - 2\gamma'_A/\gamma'_Z) \approx 1.2°$ (Prime designates the values for a pristine edge, $\gamma'_A \approx 1.01$ and $\gamma'_Z \approx 1.18$ eV/Å.)

Upon arriving at such a simple relationship, one is compelled to compare it with direct computations. Before turning to this, we note that the junctions between the A- and Z-domains along an arbitrary cut may add an AZ-mix energy correction $\delta$; proportional to the occurrence of A/Z junctions, it is evaluated as $\delta \cdot 4\sin(\chi)\sin(30°-\chi)/\cos(30°+\chi)$, and vanishes at $\chi = 0$ or $\chi = 30°$, as expected for basic edges. Energy can be computed at different levels of theory (see Supplementary Information), all to be compared with the eq. (1), whose derivation is not based on any particular model Hamiltonian. Fig. 2 shows the energies for A, Z and a few chiral edges (analogs of the low-index and the vicinal planes in crystals) computed directly with classical forces or with density functional approximations. The data of all four methods follow the theoretical curves very closely, with small and always-negative AZ-mix corrections in the range of 10 meV/Å. Moreover, a few independent calculations also fit well.[16-18]

The logic above remains unchanged if the edge is terminated by another element, but the energy definition must be augmented by subtracting the cost $\mu N$ of the terminating atoms borrowed from a reservoir of chemical potential $\mu$. If the edge is attached to a cluster of fixed size N, this constant term is of no particular interest. Often however the terminating groups are docked to the edge-atoms in one-to-one correspondence, and thus the $-\mu c$ term depends explicitly on the chiral angle. The interface energy takes form $(c_A \varepsilon_A + c_Z \varepsilon_Z) - \mu c$, and

$$\gamma(\chi) = (\sqrt{3}\gamma_A - 2\gamma_Z)\sin(\chi-30°) + (\gamma_A - 2\mu/\sqrt{3})\cos(\chi-30°) = |\gamma| \cos(\chi + C) \qquad (2)$$

In the latter, the amplitude $|\gamma|$ and phase-shift are fully defined by the values for basic A- and Z-edges, and the chemical potential $\mu$ of the terminating reactant. This analytical result can again be validated by comparison with direct ab initio computations. Fig. 2 shows good agreement (AZ-mix correction stays in the range of negative 10 meV/Å). More importantly, it reveals that the different chemistry of termination (the element x = H or Ni, and its chosen chemical potential $\mu$) does change the phase-shift C, as eq. (2) predicts.

An analytical result (2) is compact yet general. It allows one to quickly evaluate the energy for arbitrary orientation $\chi$ (especially if $\chi$ matches no rational m/n, yielding aperiodic, computationally unaffordable structures). Deriving the equilibrium shape through Wulff construction from $\gamma(\chi)$ becomes a trivial exercise.[19] The essential physics of the edge energy is all wrapped into a single parameter C: this "chemical phase" tells whether A, Z, or some intermediate edge has lowest or highest energy, and defines their ratio $\gamma_A/\gamma_Z = \cos(C+30°)/\cos(C)$. Another important characteristic is the derivative $\partial\gamma(\chi)/\partial\chi$ at the ends of the interval, $0 < \chi < 30°$, which allows one to calculate the energy of a single kink at either zigzag, $\varepsilon_Z^{kink} = {}^{\sqrt{3}}/_2\, \partial\gamma/\partial\chi|_{\chi=0} \approx \sqrt{3}\gamma_A - {}^3/_2\,\gamma_Z - {}^1/_2\mu$, or armchair edge, $\varepsilon_A^{kink} = -\,{}^1/_2\,\partial\gamma/\partial\chi|_{\chi=30°} \approx -\,{}^{\sqrt{3}}/_2\,\gamma_A + \gamma_Z$. These simple relationships are significant in reducing the great computational cost of low-symmetry kink-structures to small-unit A- or Z-edges.[20-22] Kink energies are crucial in defining the row-by-row growth of graphene or nanotubes;[12] they also define the edge stability: the rise of the $\gamma(\chi)$ curve at either end of chirality range $0 < \chi < 30°$ ensures positive kink-energy which prevents a basic edge from transforming into a vicinal. Beyond the specific useful corollaries of eq. (2), its main benefit is better seen in a big picture, resembling the "extended zone scheme" in solid state physics.

To this end, Fig. 3a shows the normalized edge energy $\gamma(\chi)/|\gamma|$ as a function of its extended argument $(\chi + C)$, where $\chi$ is a *geometrical* angle while $C$ is the *chemical* phase (determined by the chemical type of terminating element x and its chemical potential, $\mu$). For each given case, only the $\gamma_A$ and $\gamma_Z$ for basic edges need to be directly computed;[20-22] then with the proper choice of $\mu$ (e.g. $\mu = 0$ for isolated atom state, or the negative of cohesive energy for a bulk metal, etc.), the chemical phase shift is $C = \arctan[(\sqrt{3} - 2\gamma_A/\gamma_Z + \mu/\sqrt{3}\gamma_Z)/(1 - \mu/\sqrt{3}\gamma_Z)]$. In this summary plot we omit for clarity the comparison details of Fig. 2, but extend the number of examples: pristine edges (four methods), terminated by an atom-row (x = H, F, Co, Cu, Fe, Ni) or a 2D-monolayer of Ni. We first note the variability between the terminating elements, when A-edge is preferred for some, while Z-edge has lower energy with the others. We also note how, even for a given element, the change of its source (feedstock) chemical potential alters the phase C in a broad range. Interestingly, all termination types divide formally in two families, marked by different colors: if $\sqrt{3}\gamma_A < 2\gamma_Z$ then the phase C varies from -30° up to 150° (blue), while if $\sqrt{3}\gamma_A > 2\gamma_Z$ then the phase C varies from -30° down to -210° (red), upon the increase of chemical potential $\mu$. Accordingly, the 30°-wide chirality window (light blue) slides along the sinusoid, defining the edge energy behavior.

The above analysis gives the energies of graphene edges, from A to Z, through all intermediate chiral directions. It shows how the preferred orientation depends on termination and how it can—at least in principle—be broadly controlled by the chemical potential of the terminating species. Eq. (2) makes predicting the equilibrium shapes of graphene islets straightforward. Dependence on termination conditions suggests a variety of ways to control the shape of graphene during its growth.[9,10] This does not change the graphene "body" yet is important for the edge properties.

It cannot escape one's notice that the very same analysis has profound implications for nanotubes, where the origin of chirality and possibility of its control remain elusive in spite of its tremendous importance. The tube chirality is set at nucleation stage, when a complete cap (hemi-fullerene made up of hexagons and a required sextet of isolated pentagons) emerges from carbon atoms fluctuating on a catalyst.[11,23] Probability of fluctuations is controlled by the energy, which includes the catalyst, $sp^2$-carbon cap, and their contact along the circular edge, Fig. 3b-d, insets. Among these contributions only the latter depends on the edge type, determined by the angle $\chi$— also the chiral angle of the commencing tube. Therefore probability of different chiral types is defined by the edge energy, $\pi d\gamma(\chi)$. Since the diameter d is constrained by the size-fit with the catalyst particle, the cap curvature energy varies little,[23,24] leaving the chiral angle as the essential variable defining the probability, $P(\chi) \sim e^{-\pi d\gamma(\chi)/k_bT}$. We see that the preferred tube chirality is defined by the function in eq. (2). A number of observations follow. First, a strong energy bias in case of bare edge could be good for strict chirality choice, but the high energies in this case destabilize an open tube and disable its growth without a catalyst being attached, as is well known.[25] Attachment of foreign species mitigates the energy differences among chiralities, reducing $|\varepsilon_A - \varepsilon_Z|$ to a seemingly negligible several meV. The factor $\pi d \sim 30$ however brings the total edge energy variation back to values $> k_bT$ and therefore sufficient to discriminate among the chiral types. As Fig. 3 shows, chirality selection is fully determined by the phase C, depending in turn on the chemistry of species docked to the tube edge. Fig. 3 not only suggests the ways of broad variability of chiral bias, it also reveals potential difficulties due to the sheer mathematical form of cosine. It is easy to imagine a swap from A to Z preference by shifting the chirality-window from the downhill to the uphill side of the sinusoid (from Fig. 3d to c). It appears challenging, though, to tune the energy minimum to the middle of the 30°-window, to a chiral type, Fig. 3b: needed for this a convex $\gamma(\chi)$ is only available if -210° < C < -180°, in the domain of negative interface energies, where the growth is unlikely. We refrain from saying it is impossible, but it may require special quasi-equilibrium conditions, to favor a chiral tube edge yet not to cause its dissolution. It should also be noted that control by chemical potential $\mu$ is irrelevant if the catalyst is a fixed-size mono-elemental, as $-\mu N$ remains constant independent of $\chi$. On the other hand this tuning knob can be fully utilized if the number of terminating atoms directly correlates with the number of edge-atoms, which can be the case for binary compositions[26,27] with different affinity of the components to carbon. Recent experiments[28] corroborate this as a promising path.

A number of details can be added to the above theory, especially how the graphene is docked to a bulk substrate or how a cylindrical tube matches the catalyst particle, which imposes its own crystallinity and possibly facets. This complicates the analyses but can also reveal more ways for chirality control through the carbon-catalyst interface energetics. We realize the limitations of the present work, but believe it does capture and quantifies the principal factors, to offer a roadmap for graphene edge design and especially rational chirality control in nanotube production.

*** 

***

**Figures and Legends**

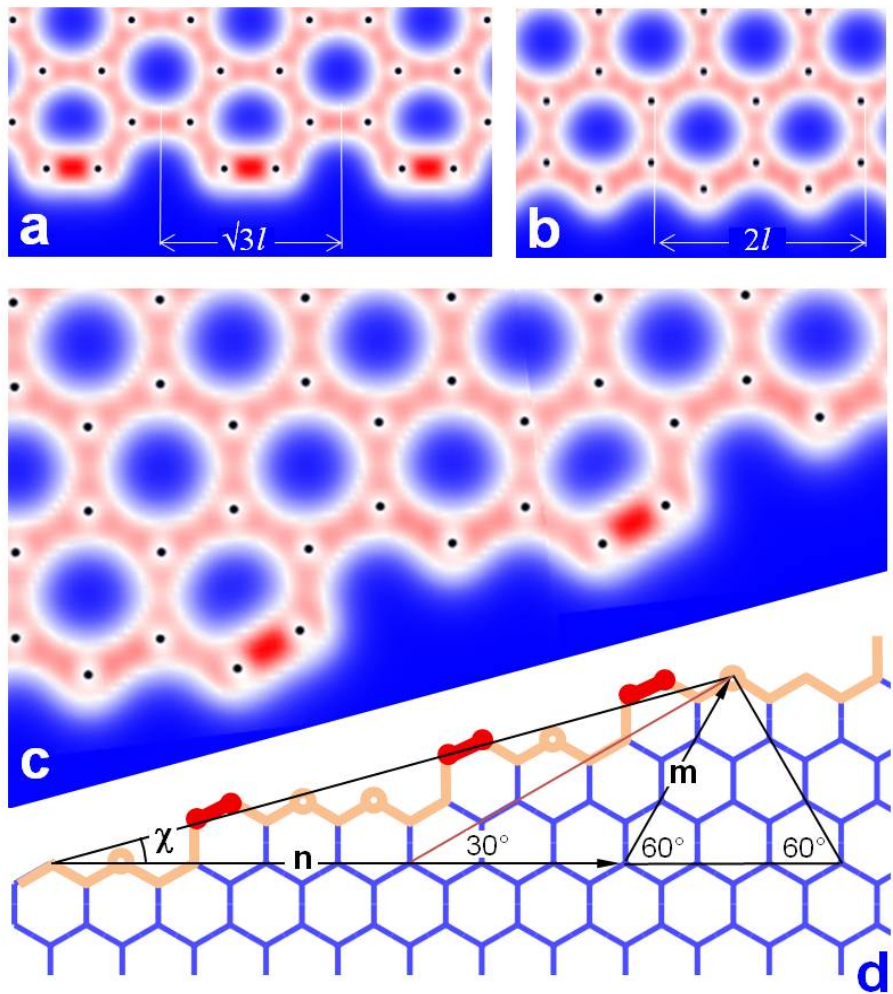

**Figure 1.** Different atomic spacing along the armchair (**a**) and zigzag (**b**) edges results in distinctly different electron density distribution, with armchair edge atoms forming shorter and stronger triple bonds. This distinction between the two types of atoms is preserved in a mixed chiral edge (**c**), as the computed electron density illustrates (from blue for zero up to red for the highest value). Schematics of the edge (**d**) along the (n, m) direction assists the atom counting: 2m A-atoms (count along the red line at 30°), and (n – m) Z-atoms (count along the horizontal black line segment). Dividing these numbers by the length $(n^2 + nm + m^2)^{1/2}$ of the edge (the diagonal on the left) yields the necessary densities, $c_A$ and $c_Z$; in this example of the (8, 3) edge there are 6 of A-atoms and 5 of Z-atoms.

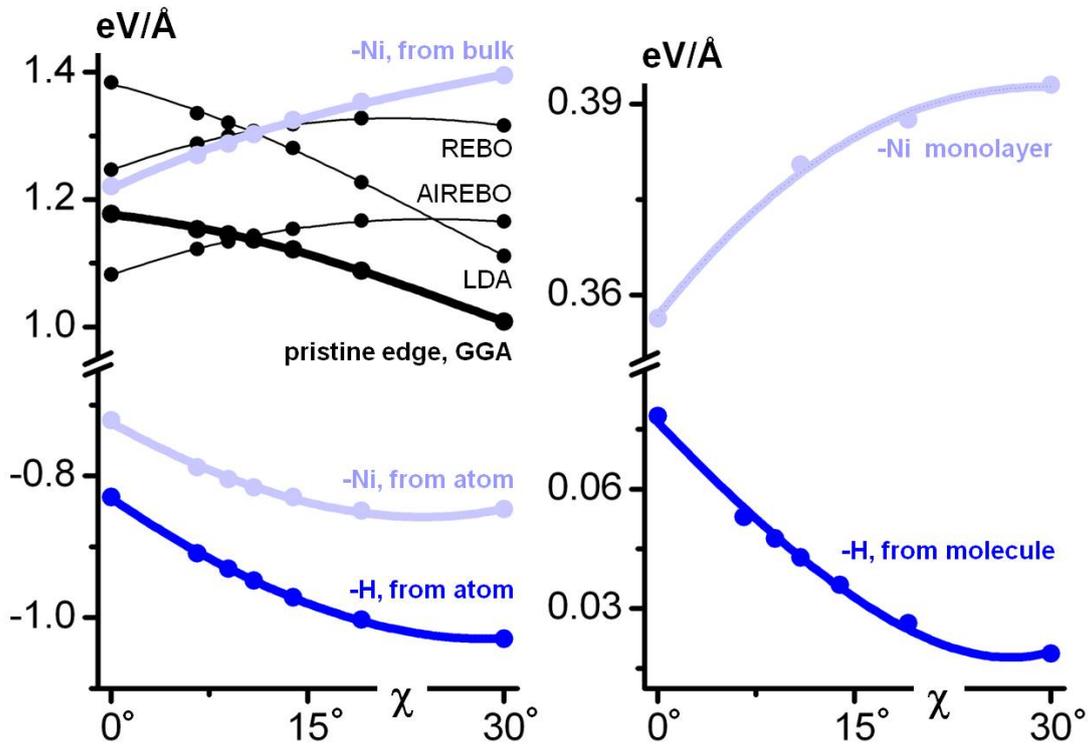

**Figure 2.** Edge energy $\gamma(\chi)/l$ as a function of chiral angle. The values are computed directly (dots) and obtained from eq. 2 (lines). Pristine edge (black) is computed with four different levels of atomistic theory, as labeled: general gradient approximation (GGA, most realistic among the four, thick line), and the local density approximation (LDA) of density functional theory, as well as classical force-fields REBO and AIREBO. Dark blue: The H-terminated edge data, with H taken either from isolated atom state or from $H_2$ molecule, to sample different chemical potential values. Light blue: Edge terminated by the Ni atom-row, either from isolated atom or from the bulk. Also in light blue is shown termination by 2-dimensional Ni atomic layer.

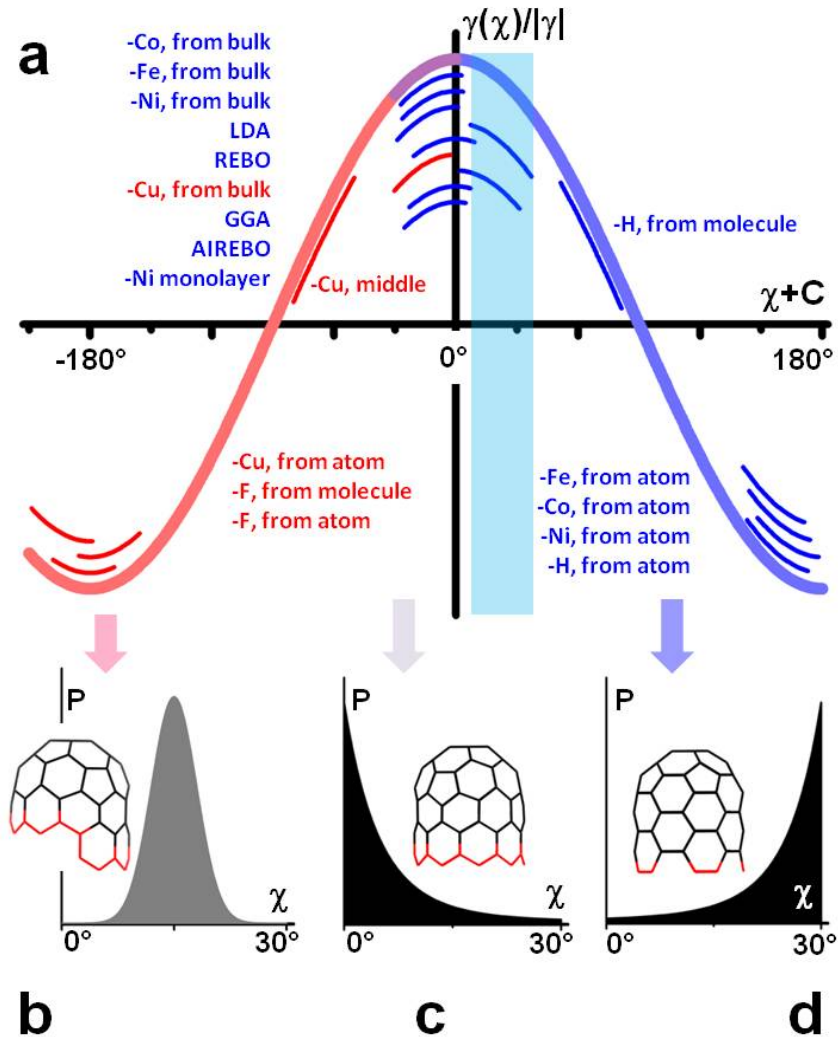

**Figure 3.** (**a**) Thick sinusoid is the normalized energy γ(χ)/|γ| versus the extended argument χ+C, r.h.s. of eq. (2). Thin segments each represent a 30°-range of chiral angle χ for each specific case of edge termination, with accordingly computed chemical phase-shifts C. The segments are labeled in the order of vertical positions, spaced to avoid overlap, for clarity. Change of the chemical potential μ can move the blue (red) segments within the blue (red) section of the sinusoid; Cu with middle μ-value is shown as example. A chiral-angle window (light blue, here placed at pristine graphene area) slides left or right according to the chemical conditions at the edge. (**b-d**) For nanotubes, the probabilities of nucleation outcome are calculated versus their edge chirality, as determined by the energy of fluctuations. The preferred chirality depends on the chemical phase C: at the left domain, a chiral tube could emerge (**b**), in the middle, the zigzag has lowest energy (**c**), and on the right, the armchair is most probable (**d**).